\documentclass[aps,prb,twocolumn,showpacs]{revtex4-1}
\usepackage{graphics}
\usepackage{epsfig}
\usepackage{epsf,epic}
\usepackage{color}
\usepackage{epstopdf}
\usepackage{subfigure}
\usepackage{amsmath}
\usepackage{amssymb}
\usepackage{amsfonts}
\usepackage{pstricks}
\usepackage{bm}
\usepackage{dcolumn}% Align table columns on decimal point
\newcommand{\etal}{\textit{et al.\ }}

\begin{document}
\title{Instability of the layered orthorhombic post-perovskite phase of SrTiO$_3$ and other candidate orthorhombic phases under pressure}
\author{Churna Bhandari}
\affiliation{Department of Physics and Astronomy,
  University of Missouri, Columbia, MO 65211 USA}
\author{Walter R. L. Lambrecht}
\affiliation{Department of Physics, Case Western Reserve University,
  Cleveland, OH 44106-7079}
\begin{abstract}
  While the tetragonal antiferro-electrically distorted (AFD) phase with
  space group $I4/mcm$ is well known for SrTiO$_3$ to occur below 105 K, 
  there are also some hints in the literature of an orthorhombic phase,
  either at the lower temperature or at high pressure. 
  A previously proposed orthorhombic layered structure of SrTiO$_3$,
  known as the post-perovskite or CaIrO$_3$ structure with space group $Cmcm$  
  is shown to have significantly higher energy than the cubic or
  tetragonal phase and to have its minimum volume at larger volume
  than cubic perovskite. The $Cmcm$ structure is thus ruled out.
  We also study an alternative
  $Pnma$ phase obtained by two octahedral rotations about different axes.
  This phase is found to have slightly lower energy than the $I4/mcm$ phase
  in spite of the fact that its parent, in-phase tilted $P4/mbm$ phase is
  not found to occur.  Our calculated enthalpies of formation
  show that the $I4/mcm$ phase occurs at slightly higher volume than the cubic
  phase and has a negative transition pressure relative to the cubic phase, which suggests that it does not correspond to the high-pressure tetragonal phase. The enthalpy of the $Pnma$ phase is almost indistinguishable from the
  $I4/mcm$ phase. Alternative ferro-electric tetragonal and orthorhombic
  structures previously suggested in literature are discussed.
\end{abstract}
\maketitle

\section{Introduction}
The phase transitions in the
perovskite material SrTiO$_3$ (STO) have been studied extensively since the
1960s.\cite{Rimai62,Rupprecht62,Bell63,Frederikse64,Lytle64}
The perovskite structure is quite versatile and allows for 
several soft-phonon mode related phase transitions. While BaTiO$_3$,
a classical ferroelectric (FE) undergoes a series of phase transitions in which
the Ti atom is displaced inside its surrounding oxygen octahedron, SrTiO$_3$
exhibits an anti-ferroelectric distortion (AFD) at 105 K consisting of a
rotation of the TiO$_6$ octahedra about one of the cubic axes. 
This leads to a tetragonal phase $I4/mcm$.\cite{Fleury68,Shirane69}
These two types of distortions are based on a soft-phonon instability
at the zone center ($\Gamma$) and at the $R$-point of the cubic Brillouin zone
respectively. 
In the first-principles based Monte-Carlo simulations model allowing for
both AFD and FE distortions and their coupling to strain, Zhong
and Vanderbilt\cite{Zhong95} predicted (at atmospheric pressure)
a second transition at about 70 K
to another tetragonal phase $I4cm$ exhibiting both types of distortions
and eventually at the lower temperature a transition to a monoclinic structure.
Under pressure, an even more complex phase diagram was predicted by these
simulations, including an orthorhombic phase and a rhombohedral phase.
Experimentally, the  lower temperature transitions have not been observed.
The absence of these transitions is now well established
to be due to the suppression of the transitions by
quantum fluctuations.\cite{Muller79,Muller91,Viana94,Zhong96} 
The enormous increase in STO dielectric response below about 40 K
has been associated with a quantum para-electric phase and possibly
the existence of some new coherent quantum phase.\cite{Muller91}
%: Fig 1
\begin{figure}
\includegraphics[width=9cm]{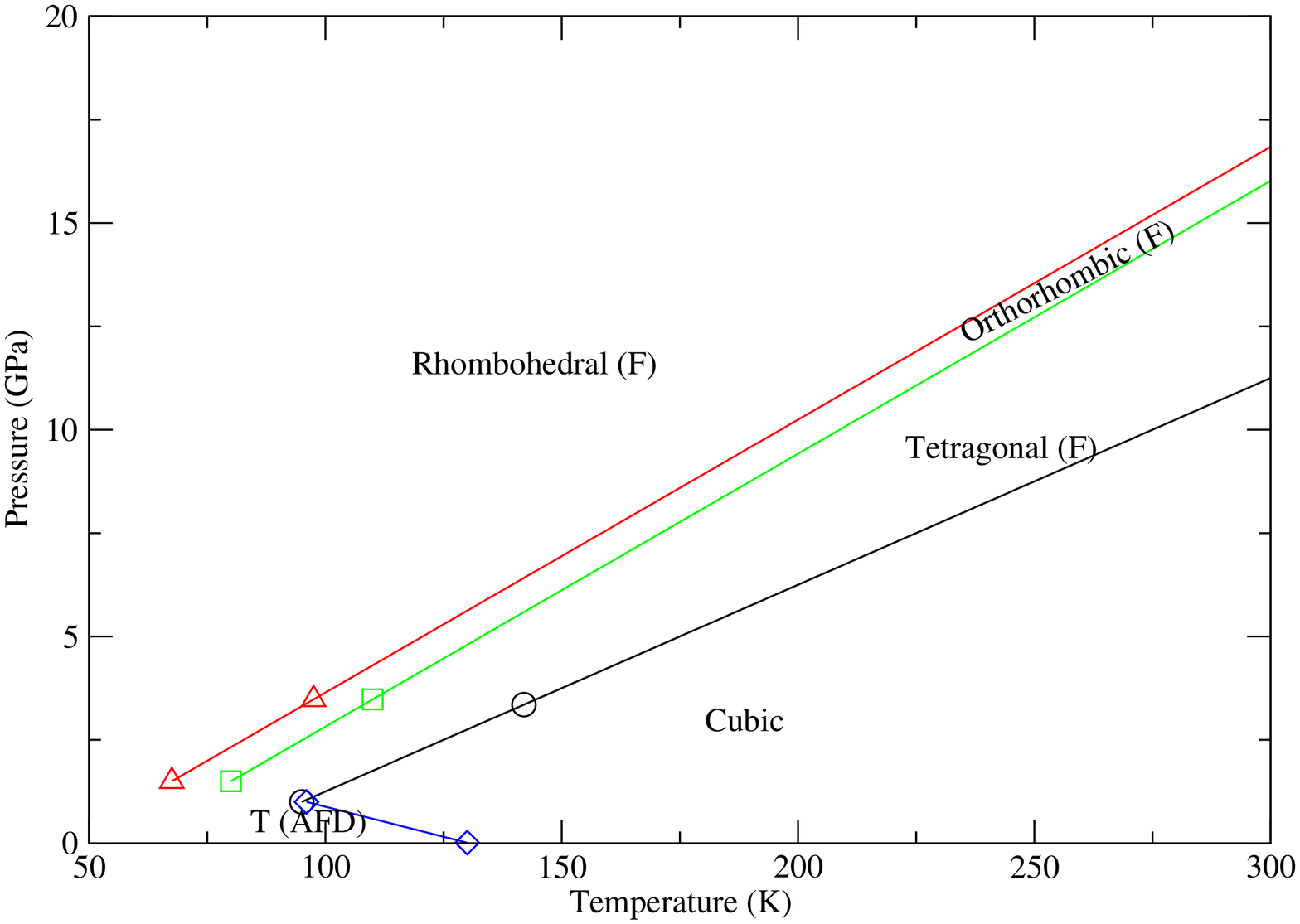}
\caption{(Color online) Extrapolated $P-T$ phase diagram of SrTiO$_3$, the data points indicated by symbols are taken from Zhong and Vanderbilt \cite{Zhong95} with their calculated
  pressures shifted from 5.5 to 0 GPa, the lines are extrapolations.\label{vanphase}}
\end{figure}
Early work by Lytle \cite{Lytle64} hinted at an orthorhombic phase at the
lower temperature but was neither confirmed nor the structure
fully determined. On the other hand, as function
of pressure, two transitions were observed, one at 5 GPa and a second
at 14-15 GPa.\cite{Ishidate88,Ishidate92,Cabaret07,Grzechnik97}
The first transition corresponds
again to the transition from cubic to a tetragonal phase.
However, it is not entirely clear whether this is the AFD $I4/mcm$
or a FE phase. According to Zhong and Vanderbilt's calculations,\cite{Zhong95}
one would rather expect a FE phase with spacegroup $P4mm$. If one
extrapolates their transition line between cubic and tetragonal FE,
to higher pressures and temperatures, a transition to this phase
is expected around 10 GPa at 300K followed by
a transition to an orthorhombic phase at slightly higher pressure
as shown in Fig. \ref{vanphase}.
This is somewhat higher
than the transition pressure observed of the cubic tetragonal transition
but clearly this extrapolation is only very approximate. 
Their calculated pressure $P=5.4$ GPa corresponds to
actual P=0 GPa because of the LDA underestimate of the lattice constants
but this shift is already included in Fig. \ref{vanphase}.

The structure  of the phase above 14 GPa has
not yet been unambiguously determined either. Raman studies by
Grzechnik \etal\cite{Grzechnik97}
suggested an orthorhombic phase. Cabaret \etal based on simulations
of the X-ray absorption oxygen K-edge and fitted to measurements
of the latter under pressure, suggested the orthorhombic CaIrO$_3$ structure.
This phase was further studied by Hachemi \etal \cite{Hachemi010},
reporting first-principles calculations of the elastic constants in each phase.
However, they did not study the transition pressure or the total
energy of this phase. Zhong and Vanderbilt\cite{Zhong95}
also mention an orthorhombic FE 
phase occurring in a narrow band between the tetragonal FE phase
and the eventually rhombohedral FE phase occurring at high pressure and low
temperature. This phase, however, is different from the CaIrO$_3$ phase.

\begin{figure}
\includegraphics[width=6cm]{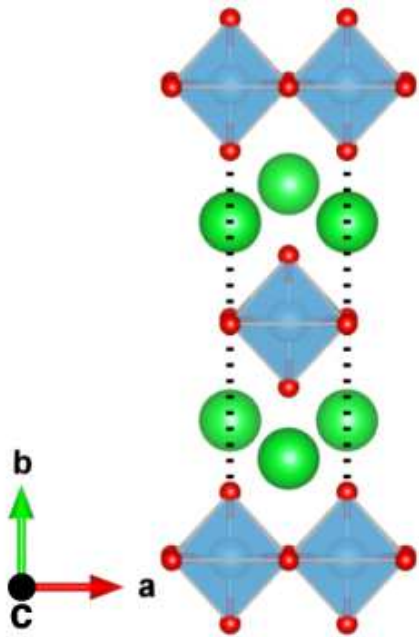}
\caption{Post-perovskite or CaIrO$_3$ structure.\label{figortho}}
\end{figure}

The structure suggested by Cabaret \etal\cite{Cabaret07} is the CaIrO$_3$
structure, with spacegroup $Cmcm$ ($D^{17}_{2h}$) reported by Rodi
and Babel.\cite{Rodi65} This structure is also known as the post-perovskite
structure and occurs for example in MgSiO$_3$ under high-pressure
in the earth's mantle.\cite{Murakami04}
Its structure is shown in Fig. \ref{figortho}. One can see that
it is quite different from the perovskite structure, requiring significant
re-bonding rather than simple soft-phonon mode distortions. It consists
of 2D layers of edge-sharing octahedra intercalated with Sr ions.
This is interesting from the point of view that from such layered structures,
it may be possible to extract atomically thin 2D nanosheets  by
exfoliation, for example by inserting larger organic ions replacing the
Sr atoms and thereby de-bonding the layers.  

%: Fig2 

Motivated by the above interest in the possibility of a
layered 2D post-perovskite orthorhombic structure of SrTiO$_3$,
we here present a study of its stability relative to the known
structures. Its  electronic structure at the quasiparticle self-consistent
$GW$ level was already reported in a separate study.\cite{Bhandaristogw}
As part of the present study, we also
calculate the energy-volume relation in tetragonal and cubic perovskite
structures. 

%: Pnma orthorhombic structure

Another possible route to an orthorhombic structure is considered.
The AFD $I4/mcm$ structure can be described as an out-of-phase
rotation of successive TiO$_6$ octahedra in the ${\bf c}$-direction,
where ${\bf c}$ is the 4-fold rotation axis. In  other words, along
these axes, the rotation is alternating clockwise and counter-clockwise.
It belongs to the Glazer rotation system, $a^0a^0c^-$.
We may alternatively consider
the in-phase rotation $a^0a^0c^+$, which corresponds to the space group
$P4/mbm$ or a soft-phonon at the $M$-point. While it was previously
found not to have an energy minimum at finite angle,\cite{NaSai2000}
it may still
be viewed as  a starting point for further octahedral rotations. 
From this tetragonal
phase, we arrive at an orthorhombic phase $Pnma$ by rotating about
a second axis, with Glazer system $a^+b^-b^-$. This would be accompanied by
motions of the Sr atoms as well. This phase occurs for example for
NaOsO$_3$, where it is called the G-type as well as in halide
perovskites, like CsSnI$_3$, where it is called the $\gamma$-phase.\cite{Lingyi14}

Recently, there has been renewed interest in SrTiO$_3$ phase transitions
because they also manifest themselves in transport properties at
the two-dimensional electron gas formed between STO and LaAlO$_3$ films
grown on top of STO substrates. The study by Schoofs \etal\cite{Schoofs12}
shows anomalies in transport at two temperatures and could also
be taken as an indication of two phase transitions. However, this may
also be related to thin film effects on the transition temperatures.
\section{Method}
We used the density functional theory\cite{Kohn-Sham} as implemented 
in the Vienna {\sl Ab initio } Simulation Package (VASP)\cite{KressePRB93,KressePRB96,KressePRB99} within the PBE exchange and correlation functional.\cite{PBEPRB86} Additionally, the non-local correlation van der Waals vdW-DF functional is used which predicts the structural parameters of weakly interacting 2D layered materials more accurately compared
to the local density (LDA) or generalized gradient
approximations (GGA). There exist various versions of vdw-DF depending upon the choice of the exchange functional used.\cite{DionPRL04,RomPRL09,JiriPCM010,KlimePRB011} In the present work we used the vdW-DF2\cite{LeePRB010} functional which is calculated within the semi-local exchange functional PW86 using the vdW-DF2 correlation kernel. For the Brillouin zone sampling, we used $8\times8\times8$ for cubic, $8\times8\times6$ for the tetragonal and orthorhombic structure respectively. The plane wave energy cut-off of 600 eV was enough for the convergence of plane wave basis-sets for all structures. The atomic positions were relaxed with the total energy converged with the precision of $10^{-6}$ eV and the force converged with the precision of 5 meV/\AA

\section{Results}
\subsection{Stability of STO under rotation}
\begin{figure}
\includegraphics[scale=0.45]{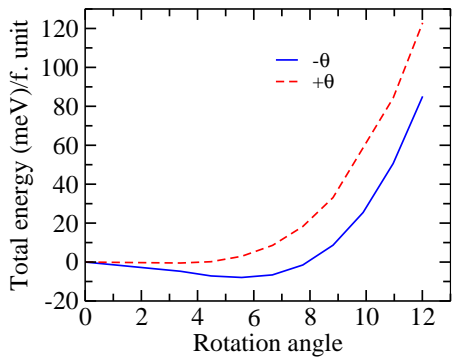}
\caption{(Color online) Total energy per formula unit as a function of octahedral rotation angle $\theta$ with fully relaxed volume and $c/a$ ratio for a given rotation angle. The dashed line corresponds to the in-phase rotation
  or $P4/mbm$ structure, the solid line corresponds to the out-of-phase or
  $I4/mcm$ structure.  The total energy is minimum for $\theta_{\rm min}=5.57^{\circ}$ for the out of-phase $I4/mcm$ phase.\label{rotang}}
\end{figure}

We first study the stability of STO under two types of AFD octahedral rotations: first the out of phase which occurs at low temperature and second the in-phase rotation which is a hypothetical structure. Fig. \ref{rotang} shows the comparison of total energy per formula unit as a function of octahedral rotation angle $\theta$. Here $+\theta$ corresponds to the rotation of adjacent octahedra in same direction along the c-axis while $-\theta$ corresponds to the opposite. Fig. \ref{rotang} shows no minimum for $\theta\ne0$ for the in-phase case. Nonetheless the minimum is very flat  (it extends over a large range of angles
between about $\pm5^\circ$) and indicates strong anharmonicity. 
On the other hand for the out of phase distortion, the total energy shows a  global minimum at a finite rotation of $\theta=5.57^{\circ}$.
These results agree with the findings of Sai and
Vanderbilt.\cite{NaSai2000}

%: Rotation Fig 3
\subsection{Energy-volume and enthalpy {\em vs.} pressure}
%:Fig 4
\begin{figure}
  \includegraphics[scale=0.4]{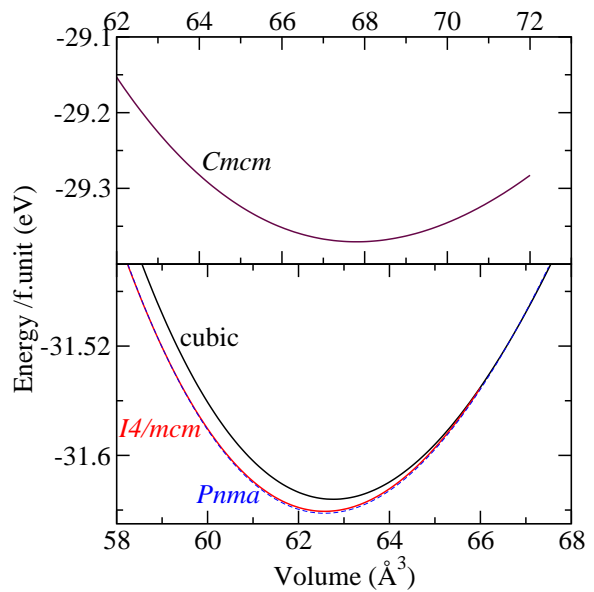}
\caption{(Color online) Pressure volume curve for cubic, tetragonal ($I4/mcm$),  orthorhombic ($Cmcm$) and ($Pnma$) structures with the PBE van der Waals functional obtained by fitting Murnaghan equation of state. \label{allphase}}
\end{figure}

%:Table I 
\begin{table}
\caption{Computed (per formula unit) minimum volume (V$_{\rm min}$), minimum energy (E$_{\rm min}$), bulk modulus (B) and derivative of bulk modulus (B$^{'}$) fitted using Murnaghan equation for all phases for pressure $p=0$ GPa and changes in volume (V) at the critical pressure $p_c$ for the cubic-$I4/mcm$ transition.}
\begin{ruledtabular}  
\begin{tabular}{lcccc}
 $p=0$&Cubic  &      $Cmcm$  &  $I4/mcm$ &   $Pnma$\\
  \hline
$\Delta E_{\rm min}$ (eV/f.u.) & 0 & 2.260  & -0.008 & -0.010  \\
%E$_{\rm min}$ & -31.632   & -29.37  &   -31.641  &  -31.642\\
V$_{\rm min}$ (\AA$^3$)   & 62.76  &     67.80 &      62.57 &     62.57\\
B (GPa)   & 172.24 &     119.933 &     169.58  &   170.82\\
B$^{'}$   & 4.405 &      4.170 &       4.32 &      4.45\\
\hline
$p_c$=-6.4 &  & & &\\ 
V&65.22 &- &65.06 &-\\
\end{tabular}
\end{ruledtabular}
\label{tab1}
\end{table}

%: Ends here
In this section, we discuss the stability of various phases  under
hydrostatic pressure. For more precise comparison, we used the van der Waals DFT which is known to predict more accurately the structural parameters for 2D like materials. In particular, the orthorhombic $Cmcm$  structure has a 2D-layered character. Fig. \ref{allphase} shows the total energy as a function of volume for four different crystal structures. The energy of the $Cmcm$ structure is much higher than the other structures which already makes its existence rather implausible.
The difference between the minima of the other phases and the $Cmcm$ structure
is 2.26 eV per formula unit. Secondly, its minimum
volume occurs at higher volume near 68 \AA$^3$, compared to $\sim 63$ \AA$^3$
for the other phases. Thus it is not likely reached by high-pressure.
We discuss this in more detail using the enthalpy {\em vs.} pressure calculations. 
On the other hand, the lower panel of Fig. \ref{allphase} shows that the energy of
the $Pnma$ phase
is very close to that of the tetragonal $I4/mcm$ phase both being lower
than the cubic phase. In fact, we find the $Pnma$ phase to have the lowest
energy but the energy difference is within the uncertainty of the calculations.

%: Churna added Table \tab1
Table \ref{tab1} shows the equilibrium volume, bulk modulus, the pressure derivative of the bulk modulus, and total energy difference relative
  to the cubic phase per formula unit. The total energy {\em vs.} volume is fitted by using the Murnaghan equation of state
\begin{equation}
E(V) = E_0 + \frac{VB}{ B'}\bigg(\frac{(V_0/V)^{B'} }{B' - 1}+ 1\bigg) - \frac{V_0B}{(B' - 1)}
\end{equation}
where $E_0$, $B$, $B^{'}$, and $V_0$ are the total energy, bulk modulus, derivative of bulk modulus, and volume at equilibrium.
The calculated bulk modulus of the cubic structure is very close to the experimental value 174 GPa.\cite{BellPR63} The bulk modulus of the $I4/mcm$ and $Pnma$ are close to that of the cubic structure, but for
the orthorhombic $Cmcm$ structure it is much smaller. The minimum volume obtained for $p=0$ GPa is lowest for the tetragonal $I4/mcm$ and  $Pnma$ structures.
This indicates that the transition pressure between cubic and the $I4/mcm$ phase
(or $Pnma$) would occur at larger volume and thus negative pressure, as is confirmed
by our enthalpy calculations, shown below. 
We also computed the volumes of the cubic and tetragonal structures at
the critical pressure $p_c=-6.4$ GPa. They occur at about 65 \AA$^3$ with
a change in volume of only 0.2 \% between the two phases and
close to the intersection of the two energy curves as can be seen in
Fig. \ref{allphase}
%:Fig 5
\begin{figure}
\subfigure[ ]{\includegraphics[scale=0.325]  {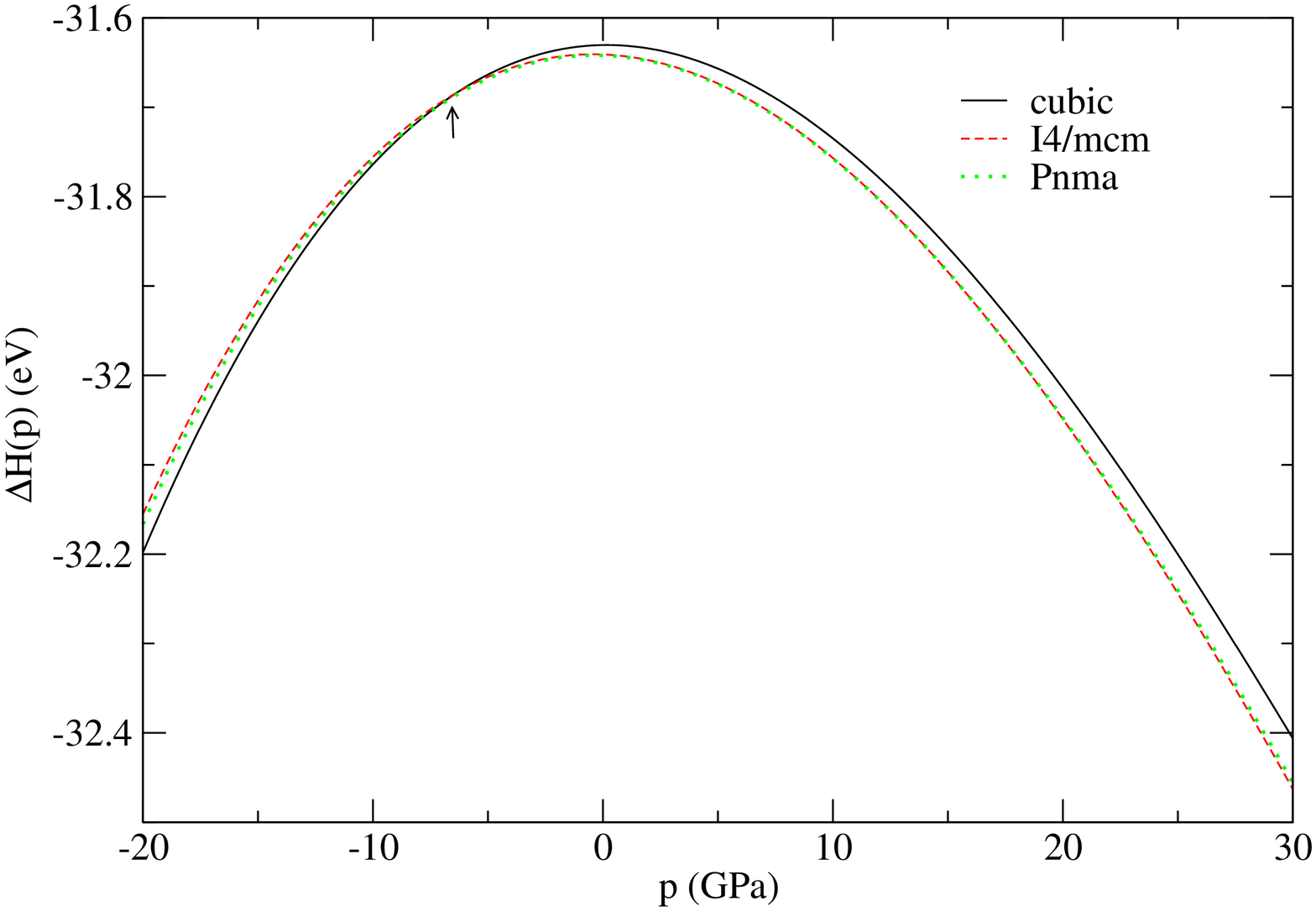}}
\subfiglabelskip 0.8em
 \subfigbottomskip 20pt
 \subfigure[ ]{ \includegraphics[scale=0.325]{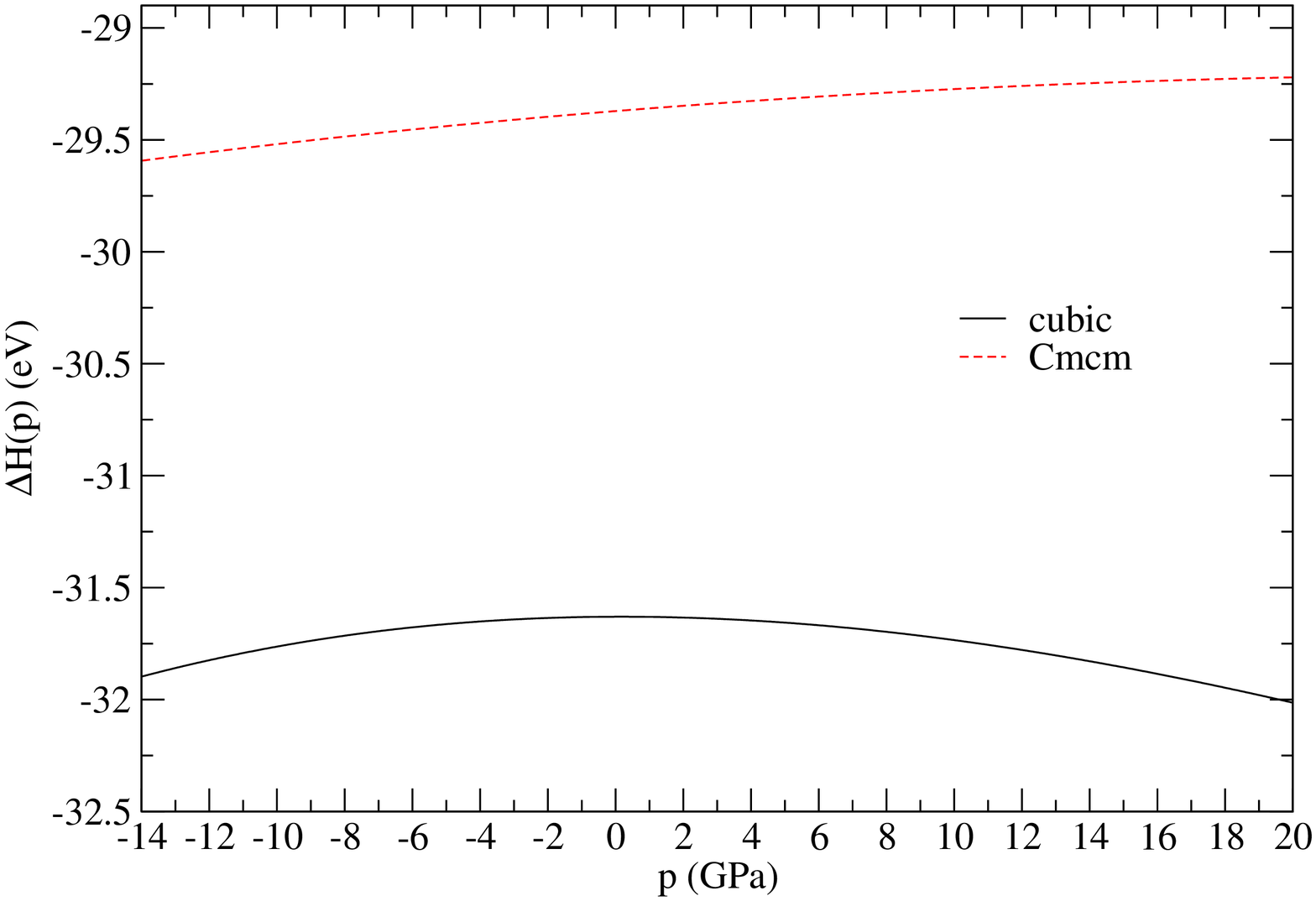}}
\caption{(Color online) Enthalpy as function of pressure
  for (a) cubic, $I4/mcm$, $Pnma$ (b) cubic and $Cmcm$ phases.
  \label{figenthalpy}}
  \end{figure}
  %:Fig 5
%\begin{figure}

Next, in Fig. \ref{figenthalpy}
we show the enthalpy changes $\Delta H=E+p\Delta V$ for the various
phases to determine the possible transition pressures given by intersections
of the enthalpies. 
We use $\Delta V=V-V_0$ with $V_0$ the equilibrium volume of the cubic phase.
This just subtracts the same $pV_0$ term for all phases and allows to
see their differences more clearly.  
We find that the cubic and $I4/mcm$ phases have crossing enthalpies
at $p_t=-6.4$ GPa. The curve for the $Pnma$ and $I4/mcm$ are indistinguishable.
However, for the cubic and $Cmcm$ phases, there is clearly no possibility
of a transition at all. 
This further rules out the possibility of the post-perovskite as high pressure
phase. 

\section{Discussion and Conclusions}
In this paper, we have studied two candidate orthorhombic structures,
the post-perovskite structure and a $Pnma$ structure. We found the
post-perovskite $Cmcm$ structure,
which has been suggested before in the literature
to correspond to a high-pressure phase of SrTiO$_3$ above 14 GPa, to
be clearly unstable. It has a higher volume and much higher total energy
than the cubic phase by about 2.26 eV/formula unit and no high-pressure
transition to this phase is possible. So, this phase can definitively
be ruled out. 
On the other hand, the $Pnma$ phase corresponding 
to a double tilt of the octahedra about two axes
and closely related to the in-phase single axis tetragonal tilted phase $P4/mbm$
was found to have slightly lower energy than
the known tetragonal $I4/mcm$  phase in spite of the fact that
an in-phase tetragonal tilt $P4/mbm$ is not favored. The energy difference
between the $Pnma$ 
and the $I4/mcm$ phase, however, is so small it may be considered to be
within the error bar of the calculations. 
The transition pressure
from the cubic phase to the $I4/mcm$ phase was calculated to be $-6.4$ GPa. The
negative value indicates that a transition from cubic to $I4/mcm$
is not a possible high-pressure transition.  The tetragonal phase found
under high pressure must thus be a different tetragonal phase than the
one obtained at low temperature.  This is consistent with 
a more complex
temperature-phase diagram, as predicted by Zhong and Vanderbilt,\cite{Zhong95} based on
Monte Carlo simulations of an effective Hamiltonian model with
degrees of freedom allowing for both ferro-electric and
antiferro-electric instabilities and parameters adjusted to
density functional calculations. The latter suggests 
a different ferro-electric tetragonal phase $P4mm$ is expected at
room temperature and at pressures above 10 GPa pressure.
None of the models thus far fully satisfactorily explains the sequence of
the two observed pressure-induced phase transitions at room temperature.
Unfortunately,
a fully first-principles molecular dynamic simulation at finite temperature and
pressure is beyond the scope of this work. Additional experimental
work at determining the precise phases occurring at high pressure appears
to be needed to resolve the question.

\acknowledgements{This work  was supported by the U.S. Air Force Office
  of Scientific Research under grant No. FA9550-18-1-0030.}

\bibliography{srto,lmto,dft,VASP}

%merlin.mbs apsrev4-1.bst 2010-07-25 4.21a (PWD, AO, DPC) hacked
%Control: key (0)
%Control: author (8) initials jnrlst
%Control: editor formatted (1) identically to author
%Control: production of article title (-1) disabled
%Control: page (0) single
%Control: year (1) truncated
%Control: production of eprint (0) enabled
\begin{thebibliography}{34}%
\makeatletter
\providecommand \@ifxundefined [1]{%
 \@ifx{#1\undefined}
}%
\providecommand \@ifnum [1]{%
 \ifnum #1\expandafter \@firstoftwo
 \else \expandafter \@secondoftwo
 \fi
}%
\providecommand \@ifx [1]{%
 \ifx #1\expandafter \@firstoftwo
 \else \expandafter \@secondoftwo
 \fi
}%
\providecommand \natexlab [1]{#1}%
\providecommand \enquote  [1]{``#1''}%
\providecommand \bibnamefont  [1]{#1}%
\providecommand \bibfnamefont [1]{#1}%
\providecommand \citenamefont [1]{#1}%
\providecommand \href@noop [0]{\@secondoftwo}%
\providecommand \href [0]{\begingroup \@sanitize@url \@href}%
\providecommand \@href[1]{\@@startlink{#1}\@@href}%
\providecommand \@@href[1]{\endgroup#1\@@endlink}%
\providecommand \@sanitize@url [0]{\catcode `\\12\catcode `\$12\catcode
  `\&12\catcode `\#12\catcode `\^12\catcode `\_12\catcode `\%12\relax}%
\providecommand \@@startlink[1]{}%
\providecommand \@@endlink[0]{}%
\providecommand \url  [0]{\begingroup\@sanitize@url \@url }%
\providecommand \@url [1]{\endgroup\@href {#1}{\urlprefix }}%
\providecommand \urlprefix  [0]{URL }%
\providecommand \Eprint [0]{\href }%
\providecommand \doibase [0]{http://dx.doi.org/}%
\providecommand \selectlanguage [0]{\@gobble}%
\providecommand \bibinfo  [0]{\@secondoftwo}%
\providecommand \bibfield  [0]{\@secondoftwo}%
\providecommand \translation [1]{[#1]}%
\providecommand \BibitemOpen [0]{}%
\providecommand \bibitemStop [0]{}%
\providecommand \bibitemNoStop [0]{.\EOS\space}%
\providecommand \EOS [0]{\spacefactor3000\relax}%
\providecommand \BibitemShut  [1]{\csname bibitem#1\endcsname}%
\let\auto@bib@innerbib\@empty
%</preamble>
\bibitem [{\citenamefont {Rimai}\ and\ \citenamefont {deMars}(1962)}]{Rimai62}%
  \BibitemOpen
  \bibfield  {author} {\bibinfo {author} {\bibfnamefont {L.}~\bibnamefont
  {Rimai}}\ and\ \bibinfo {author} {\bibfnamefont {G.~A.}\ \bibnamefont
  {deMars}},\ }\href {\doibase 10.1103/PhysRev.127.702} {\bibfield  {journal}
  {\bibinfo  {journal} {Phys. Rev.}\ }\textbf {\bibinfo {volume} {127}},\
  \bibinfo {pages} {702} (\bibinfo {year} {1962})}\BibitemShut {NoStop}%
\bibitem [{\citenamefont {Rupprecht}\ and\ \citenamefont
  {Bell}(1962)}]{Rupprecht62}%
  \BibitemOpen
  \bibfield  {author} {\bibinfo {author} {\bibfnamefont {G.}~\bibnamefont
  {Rupprecht}}\ and\ \bibinfo {author} {\bibfnamefont {R.~O.}\ \bibnamefont
  {Bell}},\ }\href {\doibase 10.1103/PhysRev.125.1915} {\bibfield  {journal}
  {\bibinfo  {journal} {Phys. Rev.}\ }\textbf {\bibinfo {volume} {125}},\
  \bibinfo {pages} {1915} (\bibinfo {year} {1962})}\BibitemShut {NoStop}%
\bibitem [{\citenamefont {Bell}\ and\ \citenamefont
  {Rupprecht}(1963{\natexlab{a}})}]{Bell63}%
  \BibitemOpen
  \bibfield  {author} {\bibinfo {author} {\bibfnamefont {R.~O.}\ \bibnamefont
  {Bell}}\ and\ \bibinfo {author} {\bibfnamefont {G.}~\bibnamefont
  {Rupprecht}},\ }\href {\doibase 10.1103/PhysRev.129.90} {\bibfield  {journal}
  {\bibinfo  {journal} {Phys. Rev.}\ }\textbf {\bibinfo {volume} {129}},\
  \bibinfo {pages} {90} (\bibinfo {year} {1963}{\natexlab{a}})}\BibitemShut
  {NoStop}%
\bibitem [{\citenamefont {Frederikse}\ \emph {et~al.}(1964)\citenamefont
  {Frederikse}, \citenamefont {Thurber},\ and\ \citenamefont
  {Hosler}}]{Frederikse64}%
  \BibitemOpen
  \bibfield  {author} {\bibinfo {author} {\bibfnamefont {H.~P.~R.}\
  \bibnamefont {Frederikse}}, \bibinfo {author} {\bibfnamefont {W.~R.}\
  \bibnamefont {Thurber}}, \ and\ \bibinfo {author} {\bibfnamefont {W.~R.}\
  \bibnamefont {Hosler}},\ }\href {\doibase 10.1103/PhysRev.134.A442}
  {\bibfield  {journal} {\bibinfo  {journal} {Phys. Rev.}\ }\textbf {\bibinfo
  {volume} {134}},\ \bibinfo {pages} {A442} (\bibinfo {year}
  {1964})}\BibitemShut {NoStop}%
\bibitem [{\citenamefont {Lytle}(1964)}]{Lytle64}%
  \BibitemOpen
  \bibfield  {author} {\bibinfo {author} {\bibfnamefont {F.~W.}\ \bibnamefont
  {Lytle}},\ }\href {\doibase http://dx.doi.org/10.1063/1.1702820} {\bibfield
  {journal} {\bibinfo  {journal} {J. Appl. Phys.}\ }\textbf {\bibinfo {volume}
  {35}},\ \bibinfo {pages} {2212} (\bibinfo {year} {1964})}\BibitemShut
  {NoStop}%
\bibitem [{\citenamefont {Fleury}\ \emph {et~al.}(1968)\citenamefont {Fleury},
  \citenamefont {Scott},\ and\ \citenamefont {Worlock}}]{Fleury68}%
  \BibitemOpen
  \bibfield  {author} {\bibinfo {author} {\bibfnamefont {P.~A.}\ \bibnamefont
  {Fleury}}, \bibinfo {author} {\bibfnamefont {J.~F.}\ \bibnamefont {Scott}}, \
  and\ \bibinfo {author} {\bibfnamefont {J.~M.}\ \bibnamefont {Worlock}},\
  }\href {\doibase 10.1103/PhysRevLett.21.16} {\bibfield  {journal} {\bibinfo
  {journal} {Phys. Rev. Lett.}\ }\textbf {\bibinfo {volume} {21}},\ \bibinfo
  {pages} {16} (\bibinfo {year} {1968})}\BibitemShut {NoStop}%
\bibitem [{\citenamefont {Shirane}\ and\ \citenamefont
  {Yamada}(1969)}]{Shirane69}%
  \BibitemOpen
  \bibfield  {author} {\bibinfo {author} {\bibfnamefont {G.}~\bibnamefont
  {Shirane}}\ and\ \bibinfo {author} {\bibfnamefont {Y.}~\bibnamefont
  {Yamada}},\ }\href {\doibase 10.1103/PhysRev.177.858} {\bibfield  {journal}
  {\bibinfo  {journal} {Phys. Rev.}\ }\textbf {\bibinfo {volume} {177}},\
  \bibinfo {pages} {858} (\bibinfo {year} {1969})}\BibitemShut {NoStop}%
\bibitem [{\citenamefont {Zhong}\ and\ \citenamefont
  {Vanderbilt}(1995)}]{Zhong95}%
  \BibitemOpen
  \bibfield  {author} {\bibinfo {author} {\bibfnamefont {W.}~\bibnamefont
  {Zhong}}\ and\ \bibinfo {author} {\bibfnamefont {D.}~\bibnamefont
  {Vanderbilt}},\ }\href {\doibase 10.1103/PhysRevLett.74.2587} {\bibfield
  {journal} {\bibinfo  {journal} {Phys. Rev. Lett.}\ }\textbf {\bibinfo
  {volume} {74}},\ \bibinfo {pages} {2587} (\bibinfo {year}
  {1995})}\BibitemShut {NoStop}%
\bibitem [{\citenamefont {M\"uller}\ and\ \citenamefont
  {Burkard}(1979)}]{Muller79}%
  \BibitemOpen
  \bibfield  {author} {\bibinfo {author} {\bibfnamefont {K.~A.}\ \bibnamefont
  {M\"uller}}\ and\ \bibinfo {author} {\bibfnamefont {H.}~\bibnamefont
  {Burkard}},\ }\href {\doibase 10.1103/PhysRevB.19.3593} {\bibfield  {journal}
  {\bibinfo  {journal} {Phys. Rev. B}\ }\textbf {\bibinfo {volume} {19}},\
  \bibinfo {pages} {3593} (\bibinfo {year} {1979})}\BibitemShut {NoStop}%
\bibitem [{\citenamefont {M{\"u}ller}\ \emph {et~al.}(1991)\citenamefont
  {M{\"u}ller}, \citenamefont {Berlinger},\ and\ \citenamefont
  {Tosatti}}]{Muller91}%
  \BibitemOpen
  \bibfield  {author} {\bibinfo {author} {\bibfnamefont {K.~A.}\ \bibnamefont
  {M{\"u}ller}}, \bibinfo {author} {\bibfnamefont {W.}~\bibnamefont
  {Berlinger}}, \ and\ \bibinfo {author} {\bibfnamefont {E.}~\bibnamefont
  {Tosatti}},\ }\href {\doibase 10.1007/BF01313549} {\bibfield  {journal}
  {\bibinfo  {journal} {Zeitschrift f{\"u}r Physik B Condensed Matter}\
  }\textbf {\bibinfo {volume} {84}},\ \bibinfo {pages} {277} (\bibinfo {year}
  {1991})}\BibitemShut {NoStop}%
\bibitem [{\citenamefont {Viana}\ \emph {et~al.}(1994)\citenamefont {Viana},
  \citenamefont {Lunkenheimer}, \citenamefont {Hemberger}, \citenamefont
  {B\"ohmer},\ and\ \citenamefont {Loidl}}]{Viana94}%
  \BibitemOpen
  \bibfield  {author} {\bibinfo {author} {\bibfnamefont {R.}~\bibnamefont
  {Viana}}, \bibinfo {author} {\bibfnamefont {P.}~\bibnamefont {Lunkenheimer}},
  \bibinfo {author} {\bibfnamefont {J.}~\bibnamefont {Hemberger}}, \bibinfo
  {author} {\bibfnamefont {R.}~\bibnamefont {B\"ohmer}}, \ and\ \bibinfo
  {author} {\bibfnamefont {A.}~\bibnamefont {Loidl}},\ }\href {\doibase
  10.1103/PhysRevB.50.601} {\bibfield  {journal} {\bibinfo  {journal} {Phys.
  Rev. B}\ }\textbf {\bibinfo {volume} {50}},\ \bibinfo {pages} {601} (\bibinfo
  {year} {1994})}\BibitemShut {NoStop}%
\bibitem [{\citenamefont {Zhong}\ and\ \citenamefont
  {Vanderbilt}(1996)}]{Zhong96}%
  \BibitemOpen
  \bibfield  {author} {\bibinfo {author} {\bibfnamefont {W.}~\bibnamefont
  {Zhong}}\ and\ \bibinfo {author} {\bibfnamefont {D.}~\bibnamefont
  {Vanderbilt}},\ }\href {\doibase 10.1103/PhysRevB.53.5047} {\bibfield
  {journal} {\bibinfo  {journal} {Phys. Rev. B}\ }\textbf {\bibinfo {volume}
  {53}},\ \bibinfo {pages} {5047} (\bibinfo {year} {1996})}\BibitemShut
  {NoStop}%
\bibitem [{\citenamefont {Ishidate}\ \emph {et~al.}(1988)\citenamefont
  {Ishidate}, \citenamefont {Sasaki},\ and\ \citenamefont
  {Inoue}}]{Ishidate88}%
  \BibitemOpen
  \bibfield  {author} {\bibinfo {author} {\bibfnamefont {T.}~\bibnamefont
  {Ishidate}}, \bibinfo {author} {\bibfnamefont {S.}~\bibnamefont {Sasaki}}, \
  and\ \bibinfo {author} {\bibfnamefont {K.}~\bibnamefont {Inoue}},\ }\href
  {\doibase 10.1080/08957958808202480} {\bibfield  {journal} {\bibinfo
  {journal} {High Pressure Research}\ }\textbf {\bibinfo {volume} {1}},\
  \bibinfo {pages} {53} (\bibinfo {year} {1988})},\ \Eprint
  {http://arxiv.org/abs/http://dx.doi.org/10.1080/08957958808202480}
  {http://dx.doi.org/10.1080/08957958808202480} \BibitemShut {NoStop}%
\bibitem [{\citenamefont {Ishidate}\ and\ \citenamefont
  {Isonuma}(1992)}]{Ishidate92}%
  \BibitemOpen
  \bibfield  {author} {\bibinfo {author} {\bibfnamefont {T.}~\bibnamefont
  {Ishidate}}\ and\ \bibinfo {author} {\bibfnamefont {T.}~\bibnamefont
  {Isonuma}},\ }\href {\doibase 10.1080/00150199208015936} {\bibfield
  {journal} {\bibinfo  {journal} {Ferroelectrics}\ }\textbf {\bibinfo {volume}
  {137}},\ \bibinfo {pages} {45} (\bibinfo {year} {1992})},\ \Eprint
  {http://arxiv.org/abs/http://dx.doi.org/10.1080/00150199208015936}
  {http://dx.doi.org/10.1080/00150199208015936} \BibitemShut {NoStop}%
\bibitem [{\citenamefont {Cabaret}\ \emph {et~al.}(2007)\citenamefont
  {Cabaret}, \citenamefont {Couzinet}, \citenamefont {Flank}, \citenamefont
  {Itié}, \citenamefont {Lagarde},\ and\ \citenamefont {Polian}}]{Cabaret07}%
  \BibitemOpen
  \bibfield  {author} {\bibinfo {author} {\bibfnamefont {D.}~\bibnamefont
  {Cabaret}}, \bibinfo {author} {\bibfnamefont {B.}~\bibnamefont {Couzinet}},
  \bibinfo {author} {\bibfnamefont {A.}~\bibnamefont {Flank}}, \bibinfo
  {author} {\bibfnamefont {J.}~\bibnamefont {Itié}}, \bibinfo {author}
  {\bibfnamefont {P.}~\bibnamefont {Lagarde}}, \ and\ \bibinfo {author}
  {\bibfnamefont {A.}~\bibnamefont {Polian}},\ }\href {\doibase
  http://dx.doi.org/10.1063/1.2644447} {\bibfield  {journal} {\bibinfo
  {journal} {AIP Conference Proceedings}\ }\textbf {\bibinfo {volume} {882}},\
  \bibinfo {pages} {120} (\bibinfo {year} {2007})}\BibitemShut {NoStop}%
\bibitem [{\citenamefont {Grzechnik}\ \emph {et~al.}(1997)\citenamefont
  {Grzechnik}, \citenamefont {Wolf},\ and\ \citenamefont
  {McMillan}}]{Grzechnik97}%
  \BibitemOpen
  \bibfield  {author} {\bibinfo {author} {\bibfnamefont {A.}~\bibnamefont
  {Grzechnik}}, \bibinfo {author} {\bibfnamefont {G.~H.}\ \bibnamefont {Wolf}},
  \ and\ \bibinfo {author} {\bibfnamefont {P.~F.}\ \bibnamefont {McMillan}},\
  }\href {\doibase
  10.1002/(SICI)1097-4555(199711)28:11<885::AID-JRS179>3.0.CO;2-Z} {\bibfield
  {journal} {\bibinfo  {journal} {Journal of Raman Spectroscopy}\ }\textbf
  {\bibinfo {volume} {28}},\ \bibinfo {pages} {885} (\bibinfo {year}
  {1997})}\BibitemShut {NoStop}%
\bibitem [{\citenamefont {Hachemi}\ \emph {et~al.}(2010)\citenamefont
  {Hachemi}, \citenamefont {Hachemi}, \citenamefont {Ferhat-Hamida},\ and\
  \citenamefont {Louail}}]{Hachemi010}%
  \BibitemOpen
  \bibfield  {author} {\bibinfo {author} {\bibfnamefont {A.}~\bibnamefont
  {Hachemi}}, \bibinfo {author} {\bibfnamefont {H.}~\bibnamefont {Hachemi}},
  \bibinfo {author} {\bibfnamefont {A.}~\bibnamefont {Ferhat-Hamida}}, \ and\
  \bibinfo {author} {\bibfnamefont {L.}~\bibnamefont {Louail}},\ }\href
  {http://stacks.iop.org/1402-4896/82/i=2/a=025602} {\bibfield  {journal}
  {\bibinfo  {journal} {Physica Scripta}\ }\textbf {\bibinfo {volume} {82}},\
  \bibinfo {pages} {025602} (\bibinfo {year} {2010})}\BibitemShut {NoStop}%
\bibitem [{\citenamefont {Rodi}\ and\ \citenamefont {Babel}(1965)}]{Rodi65}%
  \BibitemOpen
  \bibfield  {author} {\bibinfo {author} {\bibfnamefont {F.}~\bibnamefont
  {Rodi}}\ and\ \bibinfo {author} {\bibfnamefont {D.}~\bibnamefont {Babel}},\
  }\href {\doibase 10.1002/zaac.19653360104} {\bibfield  {journal} {\bibinfo
  {journal} {Zeitschrift f\"ur anorganische und allgemeine Chemie}\ }\textbf
  {\bibinfo {volume} {336}},\ \bibinfo {pages} {17} (\bibinfo {year}
  {1965})}\BibitemShut {NoStop}%
\bibitem [{\citenamefont {Murakami}\ \emph {et~al.}(2004)\citenamefont
  {Murakami}, \citenamefont {Hirose}, \citenamefont {Kawamura}, \citenamefont
  {Sata},\ and\ \citenamefont {Ohishi}}]{Murakami04}%
  \BibitemOpen
  \bibfield  {author} {\bibinfo {author} {\bibfnamefont {M.}~\bibnamefont
  {Murakami}}, \bibinfo {author} {\bibfnamefont {K.}~\bibnamefont {Hirose}},
  \bibinfo {author} {\bibfnamefont {K.}~\bibnamefont {Kawamura}}, \bibinfo
  {author} {\bibfnamefont {N.}~\bibnamefont {Sata}}, \ and\ \bibinfo {author}
  {\bibfnamefont {Y.}~\bibnamefont {Ohishi}},\ }\href {\doibase
  10.1126/science.1095932} {\bibfield  {journal} {\bibinfo  {journal}
  {Science}\ }\textbf {\bibinfo {volume} {304}},\ \bibinfo {pages} {855}
  (\bibinfo {year} {2004})},\ \Eprint
  {http://arxiv.org/abs/http://science.sciencemag.org/content}
  {http://science.sciencemag.org/content} \BibitemShut {NoStop}%
\bibitem [{\citenamefont {Bhandari}\ \emph {et~al.}(2018)\citenamefont
  {Bhandari}, \citenamefont {van Schilgaarde}, \citenamefont {Kotani},\ and\
  \citenamefont {Lambrecht}}]{Bhandaristogw}%
  \BibitemOpen
  \bibfield  {author} {\bibinfo {author} {\bibfnamefont {C.}~\bibnamefont
  {Bhandari}}, \bibinfo {author} {\bibfnamefont {M.}~\bibnamefont {van
  Schilgaarde}}, \bibinfo {author} {\bibfnamefont {T.}~\bibnamefont {Kotani}},
  \ and\ \bibinfo {author} {\bibfnamefont {W.~R.~L.}\ \bibnamefont
  {Lambrecht}},\ }\href {\doibase 10.1103/PhysRevMaterials.2.013807} {\bibfield
   {journal} {\bibinfo  {journal} {Phys. Rev. Materials}\ }\textbf {\bibinfo
  {volume} {2}},\ \bibinfo {pages} {013807} (\bibinfo {year} {2018})},\ \Eprint
  {http://arxiv.org/abs/https://doi.org/10.1103/PhysRevMaterials.2.013807}
  {https://doi.org/10.1103/PhysRevMaterials.2.013807} \BibitemShut {NoStop}%
\bibitem [{\citenamefont {Sai}\ and\ \citenamefont
  {Vanderbilt}(2000)}]{NaSai2000}%
  \BibitemOpen
  \bibfield  {author} {\bibinfo {author} {\bibfnamefont {N.}~\bibnamefont
  {Sai}}\ and\ \bibinfo {author} {\bibfnamefont {D.}~\bibnamefont
  {Vanderbilt}},\ }\href {\doibase 10.1103/PhysRevB.62.13942} {\bibfield
  {journal} {\bibinfo  {journal} {Phys. Rev. B}\ }\textbf {\bibinfo {volume}
  {62}},\ \bibinfo {pages} {13942} (\bibinfo {year} {2000})}\BibitemShut
  {NoStop}%
\bibitem [{\citenamefont {Huang}\ and\ \citenamefont
  {Lambrecht}(2014)}]{Lingyi14}%
  \BibitemOpen
  \bibfield  {author} {\bibinfo {author} {\bibfnamefont {L.-y.}\ \bibnamefont
  {Huang}}\ and\ \bibinfo {author} {\bibfnamefont {W.~R.~L.}\ \bibnamefont
  {Lambrecht}},\ }\href {\doibase 10.1103/PhysRevB.90.195201} {\bibfield
  {journal} {\bibinfo  {journal} {Phys. Rev. B}\ }\textbf {\bibinfo {volume}
  {90}},\ \bibinfo {pages} {195201} (\bibinfo {year} {2014})}\BibitemShut
  {NoStop}%
\bibitem [{\citenamefont {Schoofs}\ \emph {et~al.}(2012)\citenamefont
  {Schoofs}, \citenamefont {Egilmez}, \citenamefont {Fix}, \citenamefont
  {MacManus-Driscoll},\ and\ \citenamefont {Blamire}}]{Schoofs12}%
  \BibitemOpen
  \bibfield  {author} {\bibinfo {author} {\bibfnamefont {F.}~\bibnamefont
  {Schoofs}}, \bibinfo {author} {\bibfnamefont {M.}~\bibnamefont {Egilmez}},
  \bibinfo {author} {\bibfnamefont {T.}~\bibnamefont {Fix}}, \bibinfo {author}
  {\bibfnamefont {J.~L.}\ \bibnamefont {MacManus-Driscoll}}, \ and\ \bibinfo
  {author} {\bibfnamefont {M.~G.}\ \bibnamefont {Blamire}},\ }\href {\doibase
  10.1063/1.3687706} {\bibfield  {journal} {\bibinfo  {journal} {Applied
  Physics Letters}\ }\textbf {\bibinfo {volume} {100}},\ \bibinfo {pages}
  {081601} (\bibinfo {year} {2012})},\ \Eprint
  {http://arxiv.org/abs/http://dx.doi.org/10.1063/1.3687706}
  {http://dx.doi.org/10.1063/1.3687706} \BibitemShut {NoStop}%
\bibitem [{\citenamefont {Kohn}\ and\ \citenamefont {Sham}(1965)}]{Kohn-Sham}%
  \BibitemOpen
  \bibfield  {author} {\bibinfo {author} {\bibfnamefont {W.}~\bibnamefont
  {Kohn}}\ and\ \bibinfo {author} {\bibfnamefont {L.~J.}\ \bibnamefont
  {Sham}},\ }\href {\doibase 10.1103/PhysRev.140.A1133} {\bibfield  {journal}
  {\bibinfo  {journal} {Phys. Rev.}\ }\textbf {\bibinfo {volume} {140}},\
  \bibinfo {pages} {A1133} (\bibinfo {year} {1965})}\BibitemShut {NoStop}%
\bibitem [{\citenamefont {Kresse}\ and\ \citenamefont
  {Hafner}(1993)}]{KressePRB93}%
  \BibitemOpen
  \bibfield  {author} {\bibinfo {author} {\bibfnamefont {G.}~\bibnamefont
  {Kresse}}\ and\ \bibinfo {author} {\bibfnamefont {J.}~\bibnamefont
  {Hafner}},\ }\href {\doibase 10.1103/PhysRevB.47.558} {\bibfield  {journal}
  {\bibinfo  {journal} {Phys. Rev. B}\ }\textbf {\bibinfo {volume} {47}},\
  \bibinfo {pages} {558} (\bibinfo {year} {1993})}\BibitemShut {NoStop}%
\bibitem [{\citenamefont {Kresse}\ and\ \citenamefont
  {Furthm\"uller}(1996)}]{KressePRB96}%
  \BibitemOpen
  \bibfield  {author} {\bibinfo {author} {\bibfnamefont {G.}~\bibnamefont
  {Kresse}}\ and\ \bibinfo {author} {\bibfnamefont {J.}~\bibnamefont
  {Furthm\"uller}},\ }\href {\doibase 10.1103/PhysRevB.54.11169} {\bibfield
  {journal} {\bibinfo  {journal} {Phys. Rev. B}\ }\textbf {\bibinfo {volume}
  {54}},\ \bibinfo {pages} {11169} (\bibinfo {year} {1996})}\BibitemShut
  {NoStop}%
\bibitem [{\citenamefont {Kresse}\ and\ \citenamefont
  {Joubert}(1999)}]{KressePRB99}%
  \BibitemOpen
  \bibfield  {author} {\bibinfo {author} {\bibfnamefont {G.}~\bibnamefont
  {Kresse}}\ and\ \bibinfo {author} {\bibfnamefont {D.}~\bibnamefont
  {Joubert}},\ }\href {\doibase 10.1103/PhysRevB.59.1758} {\bibfield  {journal}
  {\bibinfo  {journal} {Phys. Rev. B}\ }\textbf {\bibinfo {volume} {59}},\
  \bibinfo {pages} {1758} (\bibinfo {year} {1999})}\BibitemShut {NoStop}%
\bibitem [{\citenamefont {Perdew}\ and\ \citenamefont {Yue}(1986)}]{PBEPRB86}%
  \BibitemOpen
  \bibfield  {author} {\bibinfo {author} {\bibfnamefont {J.~P.}\ \bibnamefont
  {Perdew}}\ and\ \bibinfo {author} {\bibfnamefont {W.}~\bibnamefont {Yue}},\
  }\href {\doibase 10.1103/PhysRevB.33.8800} {\bibfield  {journal} {\bibinfo
  {journal} {Phys. Rev. B}\ }\textbf {\bibinfo {volume} {33}},\ \bibinfo
  {pages} {8800} (\bibinfo {year} {1986})}\BibitemShut {NoStop}%
\bibitem [{\citenamefont {Dion}\ \emph {et~al.}(2004)\citenamefont {Dion},
  \citenamefont {Rydberg}, \citenamefont {Schr\"oder}, \citenamefont
  {Langreth},\ and\ \citenamefont {Lundqvist}}]{DionPRL04}%
  \BibitemOpen
  \bibfield  {author} {\bibinfo {author} {\bibfnamefont {M.}~\bibnamefont
  {Dion}}, \bibinfo {author} {\bibfnamefont {H.}~\bibnamefont {Rydberg}},
  \bibinfo {author} {\bibfnamefont {E.}~\bibnamefont {Schr\"oder}}, \bibinfo
  {author} {\bibfnamefont {D.~C.}\ \bibnamefont {Langreth}}, \ and\ \bibinfo
  {author} {\bibfnamefont {B.~I.}\ \bibnamefont {Lundqvist}},\ }\href {\doibase
  10.1103/PhysRevLett.92.246401} {\bibfield  {journal} {\bibinfo  {journal}
  {Phys. Rev. Lett.}\ }\textbf {\bibinfo {volume} {92}},\ \bibinfo {pages}
  {246401} (\bibinfo {year} {2004})}\BibitemShut {NoStop}%
\bibitem [{\citenamefont {Rom\'an-P\'erez}\ and\ \citenamefont
  {Soler}(2009)}]{RomPRL09}%
  \BibitemOpen
  \bibfield  {author} {\bibinfo {author} {\bibfnamefont {G.}~\bibnamefont
  {Rom\'an-P\'erez}}\ and\ \bibinfo {author} {\bibfnamefont {J.~M.}\
  \bibnamefont {Soler}},\ }\href {\doibase 10.1103/PhysRevLett.103.096102}
  {\bibfield  {journal} {\bibinfo  {journal} {Phys. Rev. Lett.}\ }\textbf
  {\bibinfo {volume} {103}},\ \bibinfo {pages} {096102} (\bibinfo {year}
  {2009})}\BibitemShut {NoStop}%
\bibitem [{\citenamefont {Klimeš}\ \emph {et~al.}(2010)\citenamefont
  {Klimeš}, \citenamefont {Bowler},\ and\ \citenamefont
  {Michaelides}}]{JiriPCM010}%
  \BibitemOpen
  \bibfield  {author} {\bibinfo {author} {\bibfnamefont {J.}~\bibnamefont
  {Klimeš}}, \bibinfo {author} {\bibfnamefont {D.~R.}\ \bibnamefont {Bowler}},
  \ and\ \bibinfo {author} {\bibfnamefont {A.}~\bibnamefont {Michaelides}},\
  }\href {http://stacks.iop.org/0953-8984/22/i=2/a=022201} {\bibfield
  {journal} {\bibinfo  {journal} {Journal of Physics: Condensed Matter}\
  }\textbf {\bibinfo {volume} {22}},\ \bibinfo {pages} {022201} (\bibinfo
  {year} {2010})}\BibitemShut {NoStop}%
\bibitem [{\citenamefont {Klime\ifmmode~\check{s}\else \v{s}\fi{}}\ \emph
  {et~al.}(2011)\citenamefont {Klime\ifmmode~\check{s}\else \v{s}\fi{}},
  \citenamefont {Bowler},\ and\ \citenamefont {Michaelides}}]{KlimePRB011}%
  \BibitemOpen
  \bibfield  {author} {\bibinfo {author} {\bibfnamefont {J.~c.~v.}\
  \bibnamefont {Klime\ifmmode~\check{s}\else \v{s}\fi{}}}, \bibinfo {author}
  {\bibfnamefont {D.~R.}\ \bibnamefont {Bowler}}, \ and\ \bibinfo {author}
  {\bibfnamefont {A.}~\bibnamefont {Michaelides}},\ }\href {\doibase
  10.1103/PhysRevB.83.195131} {\bibfield  {journal} {\bibinfo  {journal} {Phys.
  Rev. B}\ }\textbf {\bibinfo {volume} {83}},\ \bibinfo {pages} {195131}
  (\bibinfo {year} {2011})}\BibitemShut {NoStop}%
\bibitem [{\citenamefont {Lee}\ \emph {et~al.}(2010)\citenamefont {Lee},
  \citenamefont {Murray}, \citenamefont {Kong}, \citenamefont {Lundqvist},\
  and\ \citenamefont {Langreth}}]{LeePRB010}%
  \BibitemOpen
  \bibfield  {author} {\bibinfo {author} {\bibfnamefont {K.}~\bibnamefont
  {Lee}}, \bibinfo {author} {\bibfnamefont {E.~D.}\ \bibnamefont {Murray}},
  \bibinfo {author} {\bibfnamefont {L.}~\bibnamefont {Kong}}, \bibinfo {author}
  {\bibfnamefont {B.~I.}\ \bibnamefont {Lundqvist}}, \ and\ \bibinfo {author}
  {\bibfnamefont {D.~C.}\ \bibnamefont {Langreth}},\ }\href {\doibase
  10.1103/PhysRevB.82.081101} {\bibfield  {journal} {\bibinfo  {journal} {Phys.
  Rev. B}\ }\textbf {\bibinfo {volume} {82}},\ \bibinfo {pages} {081101}
  (\bibinfo {year} {2010})}\BibitemShut {NoStop}%
\bibitem [{\citenamefont {Bell}\ and\ \citenamefont
  {Rupprecht}(1963{\natexlab{b}})}]{BellPR63}%
  \BibitemOpen
  \bibfield  {author} {\bibinfo {author} {\bibfnamefont {R.~O.}\ \bibnamefont
  {Bell}}\ and\ \bibinfo {author} {\bibfnamefont {G.}~\bibnamefont
  {Rupprecht}},\ }\href {\doibase 10.1103/PhysRev.129.90} {\bibfield  {journal}
  {\bibinfo  {journal} {Phys. Rev.}\ }\textbf {\bibinfo {volume} {129}},\
  \bibinfo {pages} {90} (\bibinfo {year} {1963}{\natexlab{b}})}\BibitemShut
  {NoStop}%
\end{thebibliography}%

\end{document}